\RequirePackage{fix-cm}
\documentclass[smallextended]{svjour3}       
\smartqed  
\usepackage{graphicx}
\usepackage{lineno,hyperref}
\usepackage{graphicx,amssymb,arydshln,subeqnarray,subfigure}
\usepackage{url}
\usepackage{amsbsy,amsmath,amsfonts,cases,bm,caption,color}
\usepackage[ruled]{algorithm2e} 
\modulolinenumbers[5]

\newcommand{\tabincell}[2]{\begin{tabular}{@{}#1@{}}#2\end{tabular}}
\begin{document}

\title{Maximum Correntropy Criterion for Robust TOA-Based Localization in NLOS Environments
}


\author{Wenxin Xiong \and Christian Schindelhauer \\\and Hing Cheung So \and Zhi Wang}


\institute{W. Xiong \at
           Department of Computer Science, University of Freiburg, Freiburg 79110, Germany \\
              \email{xiongw@informatik.uni-freiburg.de}           
           \and
           C. Schindelhauer \at
           Department of Computer Science, University of Freiburg, Freiburg 79110, Germany \\
		   \email{schindel@informatik.uni-freiburg.de} 
           \and
           H. C. So \at
           Department of Electrical Engineering, City University of Hong Kong, Hong Kong, China \\
		   \email{hcso@ee.cityu.edu.hk}
           \and
           Z. Wang \at
           State Key Laboratory of Industrial Control Technology, Zhejiang University, Hangzhou 311121, China \\
		   \email{zjuwangzhi@zju.edu.cn}
}

\date{Received: date / Accepted: date}

\maketitle

\begin{abstract}
We investigate the problem of time-of-arrival (TOA) based localization under possible non-line-of-sight (NLOS) propagation conditions. To robustify the squared-range-based location estimator, we follow the maximum correntropy criterion, essentially the Welsch $M$-estimator with a redescending influence function which behaves like $\ell_0$-minimization towards the grossly biased measurements, to derive the formulation. The half-quadratic technique is then applied to settle the resulting optimization problem in an alternating maximization (AM) manner. By construction, the major computational challenge at each AM iteration boils down to handling an easily solvable generalized trust region subproblem. It is worth noting that the implementation of our localization method requires nothing but merely the TOA-based range measurements and sensor positions as prior information. Simulation and experimental results demonstrate the competence of the presented scheme in outperforming several state-of-the-art approaches in terms of positioning accuracy, especially in scenarios where the percentage of NLOS paths is not large enough.
\end{abstract}

\section{Introduction}
\label{intro}
Source localization based on location-bearing information gathered at spatially separated sensors \cite{HCSo} plays a pivotal role in many science and engineering areas such as cellular networks \cite{JAdPR}, Internet of Things \cite{FZafari}, and wireless sensor networks \cite{FXiao}. Being perhaps the most popular measurement model, time-of-arrival (TOA) defined as the one-way travel time of the signal between the emitting source and a sensor has co-existed with numerous communication technologies for positioning ranging across ZigBee \cite{JCheon}, radio frequency identification device \cite{TFBechteler}, ultra-wideband (UWB) \cite{ARJRuiz}, and ultrasound \cite{FHoeflinger}, and will be the main focus herein.

A challenging issue in this context is that due to the obstruction of signal transmissions between the source and sensors, non-line-of-sight (NLOS) propagation is generally unavoidable in the real-world scenarios (e.g., urban canyons and indoor locales). The NLOS error in a contaminated TOA appears as a positive bias because of additional propagation delay, indicating that special attention has to be paid to alleviating its adverse impacts on positioning accuracy. While studies of TOA-based localization under NLOS conditions may date back more than one-and-a-half decades \cite{IGuvenc}, NLOS mitigation schemes subject to relatively few specific assumptions about the errors have yet only lately been investigated in the literature \cite{STomic2,IGuvenc,GWang,SZhang,STomic1,HChen2,RMVaghefi2,CHPark,WXiong,FYin1,FYin2}.

The first branch of these methods takes a so-called estimation-based strategy to alleviate the adverse impacts of NLOS conditions on positioning accuracy. For instance, as the primary contribution of \cite{GWang}, the authors propose to replace multiple NLOS bias errors by only one (viz., a balancing parameter to be estimated), based on which the effects of NLOS propagation are partially mitigated. Next, convex relaxation techniques \cite{SBoyd} including second-order cone programming (SOCP) and semidefinite programming (SDP) are employed to tackle the formulation with nonconvexity. The tactic of jointly estimating the source location and a balancing parameter is later reused in \cite{STomic2}, only the solving process thereof is organized in a two-step weighted least squares (LS) manner while the unconstrained minimization problem in each step, by construction, falls into a computationally simpler generalized trust region subproblem (GTRS) framework \cite{ABeck} and thus can be addressed exactly. Apart from them, in \cite{RMVaghefi2}, a set of bias-like terms are treated as the optimization variables in addition to those for the source position. The authors then discard the constraints between these new variables and NLOS errors, and put forward a distinct SDP estimator to eliminate the nonconvexity of the established nonlinear LS problem.

Instead of precisely setting the NLOS-error-related optimization variables, one may model the uncertainties robustly using a less sensitive worst-case criterion \cite{GWang,SZhang,STomic1,HChen2}, i.e., searching for parameters over all plausible values that have the best possible performance in the worst-case sense \cite{SBoyd}. The essence of this scheme is to exploit the predetermined upper bounds on the NLOS errors, which are more readily ascertainable compared to their distribution/statistics and the path status \cite{GWang}. Specifically, a robust SDP method built upon the S-procedure \cite{SBoyd} is developed in \cite{GWang}, whereas the approximations without leveraging S-procedure are made in \cite{SZhang} and \cite{STomic1}, finally boiling down to a robust SOCP method and a bisection-based robust GTRS solution, respectively.

Toward a complementarity between the aforementioned two categories of methodologies, a more recent work \cite{HChen2} turns to regard the NLOS error in a TOA measurement as the superposition of a balancing parameter and a new variable to which robustness is conferred. Bearing a close resemblance to \cite{GWang}, the S-procedure is followed to eliminate the maximization part of the cumbersome minimax problem, whereupon the semidefinite relaxation is conducted to yield a tractable convex program. To boost the resilience of TOA-based localization system, there are also frequently chosen options other than the worst-case formulation which are less heavily dependent on the prior knowledge of NLOS information, e.g., the recursive Bayesian approaches with robust statistics in \cite{CHPark}, model parameter determination of probability density function for the non-Gaussian distributions in \cite{FYin1,FYin2}, and robust multidimensional similarity analysis (RMDSA) in \cite{WXiong} borrowing the idea from outlier-resistant low-rank matrix completion, to name just a few.

Robust statistics based schemes usually benefit from their removal of requirements for \textit{a priori} noise/error information and, therefore, fit in perfectly with the practical localization applications. Such an assumption is in contrast to the majority of existing work, e.g. \cite{IGuvenc,GWang,SZhang,STomic1,HChen2,RMVaghefi2}, which more or less rely on the prior knowledge about noise variance/error bounds, in addition to the TOA-based range measurements and sensor positions. Motivated by its $\ell_0$-like insensitivity toward grossly biased samples and widespread use in non-Gaussian signal processing including robust low-rank tensor recovery \cite{YYang} and robust radar target localization \cite{JLiang}, the correntropy measure \cite{WLiu}, essentially a Welsch $M$-estimator based cost function, is herein utilized for achieving higher degree of resistance to the NLOS errors. The half-quadratic (HQ) theory \cite{MNikolova} is then exploited to convert the reshaped maximum correntropy criterion (MCC) estimation problem into a sequence of quadratic optimization tasks \cite{SBoyd}, after which the computationally attractive GTRS technique is applicable. It is noteworthy that our MCC-induced robustification is imposed upon the squared-range (SR) \cite{ABeck} rather than range measurement model. This, as we show in Section \ref{AD}, can make the development of the HQ algorithm more tractable. Furthermore, our localization approach does not require any extra prior information except the TOA-based range measurements and sensor positions.

The remainder of this paper is organized as follows. Section \ref{PFP} justifies our use of the noise/error mixture model and correntropy measure, and formulates the robust estimation problem. Section \ref{AD} expatiates the derivation process and important properties of the proposed algorithm. In Section \ref{NR}, numerical results are included. Finally, conclusions are drawn in Section \ref{CC}.

\section{Preliminaries and problem formulation}
\label{PFP}
Consider $L \geq d+1$ sensors and a single source in the $d$-dimensional space ($d = 2$ or $3$). Denoting the known position of the $i$th sensor and unknown source location by $\bm{x}_i \in \mathbb{R}^d$ (for $i = 1,...,L$) and $\bm{x} \in \mathbb{R}^d$, respectively, the TOA-based range measurement between the $i$th sensor and source is modeled as $r_i = {\|\bm{x} - \bm{x}_i\|}_2 + e_i$, where ${\|\cdot\|}_2$ stands for the $\ell_2$-norm, and $e_i$ is the error in the ranging observation $r_i$ under possible NLOS propagation conditions, following a mixture model of Gaussian and non-Gaussian distributions. In this mixture model, the relatively lower-level Gaussian distributed term represents the measurement noise due to thermal disturbance at the sensor, whereas the non-Gaussian counterpart stands for the NLOS bias error in the corresponding source-sensor path. Also notable is that the similar noise/error modeling schemes have been widely reported in the literature on TOA-based source localization under NLOS propagation \cite{IGuvenc}. While the recent efforts tend to perform error mitigation using as little NLOS information as possible, it is increasingly common to generalize the NLOS bias error term (i.e., one does not assume any specific non-Gaussian distribution) in the derivation of robust location estimators \cite{STomic2,GWang,SZhang,STomic1,HChen2,RMVaghefi2,WXiong}. Depending on what kind of distributions are applied to generate the NLOS errors for simulation, these studies can be classified into the exponential \cite{RMVaghefi2} and uniform \cite{STomic2,GWang,SZhang,STomic1,HChen2,WXiong} ones.

In this paper, we adopt the aforesaid robust localization setting, in which no prior knowledge about the statistics of NLOS bias errors or the error status is available to the algorithm in the problem-solving stage. By convention, the only information we assume is that the non-Gaussian error term in $e_i$ (in the NLOS scenarios) is positive and possesses the bias-like feature, namely its magnitude is much larger than that of the Gaussian random process. We simply follow the more frequently used uniform distribution to produce the non-Gaussian turbulence in $e_i$ in our computer simulations. Note that there are also other noise/error modeling strategies among the related work discussed in Section \ref{intro}, such as the Gaussian mixture of two components \cite{CHPark,FYin1,FYin2} and Gaussian-Laplace mixture \cite{FXiao}. Since both Gaussian and Laplace distributions are with infinite support, they are normally utilized for the approximations of impulsive noise rather than the positively biased NLOS errors.

\begin{figure}[!t]
	\centering
	\includegraphics[width=3.5in]{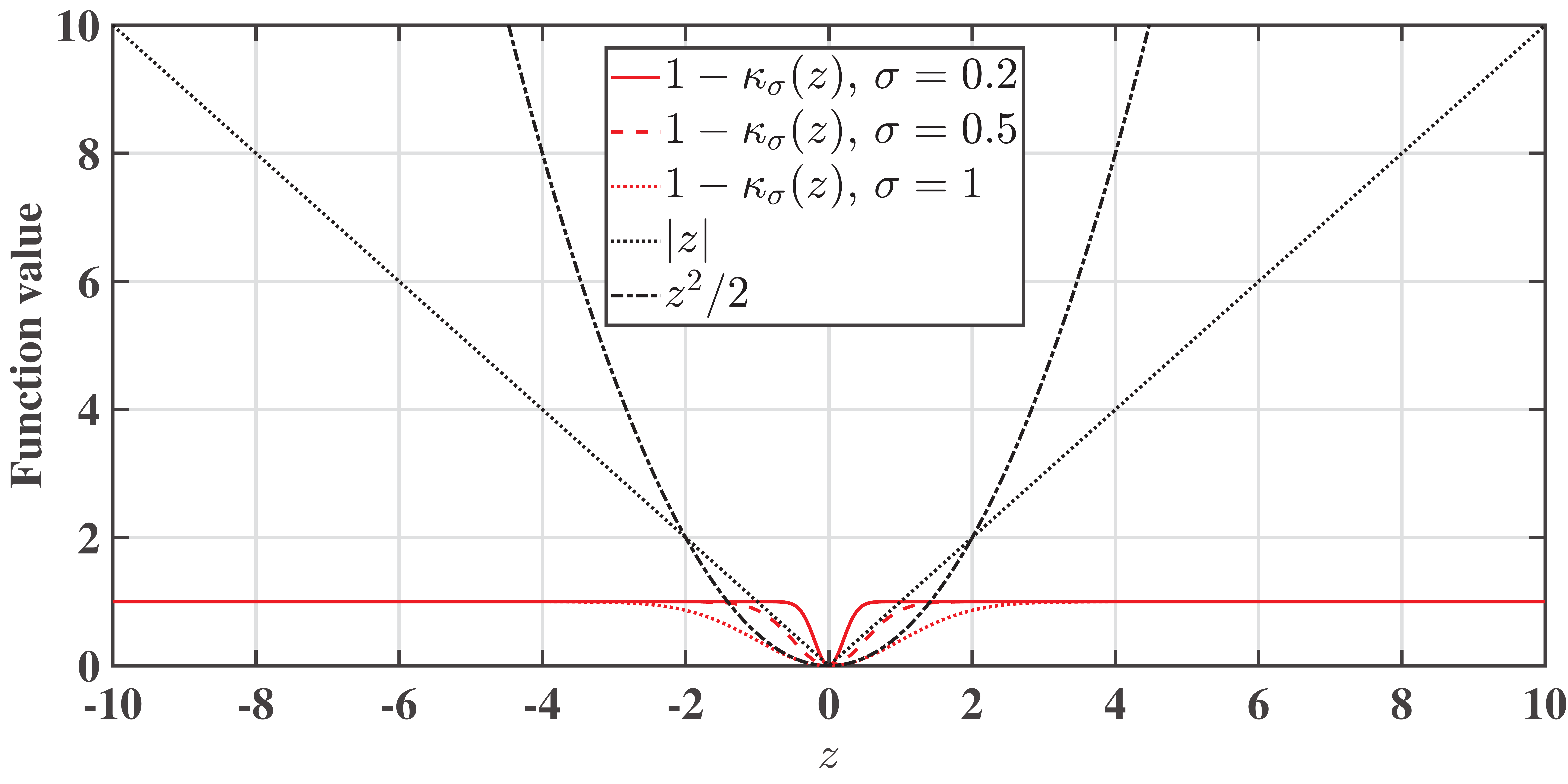}
	\caption{Comparison of different loss functions: $1-\kappa_{\sigma}(x)$, $|z|$, and $z^2/2$.} 
	\label{Losses_Comp}
\end{figure}

A local, nonlinear, and generalized similarity measure between two random variables $X$ and $Y$, known as the correntropy \cite{WLiu}, is defined as $V_{\sigma}(X,Y) = \mathbf{E} \left[\kappa_{\sigma} (X-Y)\right]$, where $\mathbf{E}\left[\cdot\right]$ denotes the expectation operator and $\kappa_{\sigma}(x)$ is the kernel function with size $\sigma$ satisfying the Mercer's theorem \cite{VVapnik}. In this paper, we fix $\kappa_{\sigma}(x)$ as the Gaussian kernel, i.e., $\kappa_{\sigma}(x) = \exp \left(-x^2/(2\sigma^2)\right)$. In the practical scenarios where only a finite amount of data $\left\{ X_i, Y_i \right\}_{i=1}^N$ is available, the sample estimator of correntropy: $\hat{V}_{N,\sigma} (X,Y) = \frac{1}{N} \sum_{i=1}^{N} \kappa_{\sigma} (X_i - Y_i)$ is used instead. The MCC aiming at maximizing the sample correntropy function, or equivalently, minimizing its decreasing function which is closely associated with the Welsch $M$-estimator, has found many applications in non-Gaussian signal processing \cite{YYang,JLiang}. Equipped with a redescending influence function, Welsch $M$-estimator is accepted to outperform not just $\ell_2$- and $\ell_1$-minimization criteria but also the Huber and Cauchy $M$-estimators in terms of outlier-robustness \cite{YYang}, while on the other side, have the advantage of being smoother than the Tukey's biweight $M$-estimator \cite{AMZoubir}. For comparative purposes, Fig. \ref{Losses_Comp} plots $|z|$, $z^2/2$, and $1-\kappa_{\sigma}(z)$ with different $\sigma$s. We observe that $1-\kappa_{\sigma}(z)$, essentially the Welsch loss, can well approximate the $\ell_2$ loss and hence be statistically quite efficient with respect to (w.r.t.) lower-level Gaussian disturbance. Oppositely, it will eventually saturate, behave like cardinality, and exhibit insensitivity to outliers as the magnitude of $z$ increases. What is more, all of its properties are controlled by the kernel size $\sigma$. These characteristics have justified our use of the correntropy measure for handling the bias-like NLOS errors.

Based on the MCC, a maximization problem is formulated as
\begin{equation}{\label{MCCTOA}}
\max_{\bm{x}} \sum_{i=1}^{L} \kappa_{\sigma}\left(r_i^2 - {\|\bm{x} - \bm{x}_i\|}_2^2\right).
\end{equation}
It should be noted that the fitting errors in (\ref{MCCTOA}) are expressed using the SR model \cite{ABeck} instead of the range-based one, i.e., $X_i - Y_i = r_i^2 - {\|\bm{x} - \bm{x}_i\|}_2^2$. As illustrated in what follows, such a treatment is crucial for a computationally simple $\bm{x}$-ascertainment step in solving (\ref{MCCTOA}).

\section{Algorithm development}
\label{AD}
The MCC-based optimization problem (\ref{MCCTOA}) is in general difficult to solve because of the severe nonconvexity. In this section, we tackle it based on the HQ reformulation and bisection-based GTRS solution.

According to the HQ theory \cite{MNikolova}, there exists a convex conjugate function $\zeta: \mathbb{R} \rightarrow \mathbb{R}$ of $\kappa_{\sigma}(x)$ so that $\kappa_{\sigma}(x) = \max_{p} \left( p \frac{x^2}{\sigma^2} - \zeta(p) \right)$, and for any fixed $x$, the maximum is attained at $p = -\kappa_{\sigma}(x)$.

By employing the HQ technique, (\ref{MCCTOA}) is reformulated as
\begin{equation}{\label{MCCTOA2}}
\max_{\check{\bm{x}}} \mathcal{A}_{\sigma}(\bm{x},\bm{p}) := \sum_{i=1}^{L} \left[ p_i \frac{{\left(r_i^2 - {\|\bm{x} - \bm{x}_i\|}_2^2\right)}^2}{\sigma^2} - \zeta(p_i) \right],
\end{equation}
where $\check{\bm{x}} = \left[ \bm{x}^T, \bm{p}^T \right]^T \in \mathbb{R}^{d+L}$ and $\bm{p} = \left[ p_1,p_2,...,p_L \right]^T \in \mathbb{R}^L$ is a vector containing the auxiliary variables. This can also be interpreted as introducing an augmented cost function $\mathcal{A}_{\sigma}$ in the enlarged parameter space $\left\{ \bm{x}, \bm{p} \right\}$. A local maximizer of (\ref{MCCTOA2}) is then calculated using the following alternating maximization (AM) procedure:
\begin{subequations}{\label{eq:case2}}
\begin{align}
\bm{p}_{\left(k+1\right)}&= \arg\max_{\bm{p}} \mathcal{A}_{\sigma}\left(\bm{x}_{\left(k\right)},\bm{p}\right) \label{pp} \\
\bm{x}_{\left(k+1\right)}&= \arg\max_{\bm{x}} \mathcal{A}_{\sigma}\left(\bm{x},\bm{p}_{\left(k+1\right)}\right) \label{px}
\end{align}
\end{subequations}
where the subscript $(\cdot)_{\left( k \right)}$ denotes the iteration index.

We can derive from the properties of convex conjugate function and simple observations that the solution of sub-problem (\ref{eq:case2}a) is
\begin{equation}{\label{s_subp1}}
\left[\bm{p}_{\left(k+1\right)}\right]_{i} = -\exp \left(-\frac{{\left(r_i^2 - {\|\bm{x}_{\left(k\right)} - \bm{x}_i\|}_2^2\right)}^2}{2 \sigma^2}\right),
\end{equation}
where $\left[ \cdot \right]_i \in \mathbb{R}$ represents the $i$th element of a vector. By ignoring the constant terms independent of the optimization variable $\bm{x}$ and rewriting the problem into a minimization form, the sub-problem (\ref{eq:case2}b) amounting to the SR-LS estimation \cite{ABeck} problem
\begin{equation*}
\min_{\bm{x}} \sum_{i=1}^{L} \left\{-\left[\bm{p}_{\left(k+1\right)}\right]_{i} {\left( {\|\bm{x} - \bm{x}_i\|}_2^2 - r_i^2 \right)}^2\right\}
\end{equation*}
can actually be transformed into a GTRS w.r.t. $\bm{y} = \left[ \bm{x}^T, \alpha \right]^T \in \mathbb{R}^{d+1}$, viz.
\begin{align}{\label{GTRS}}
\min_{\bm{y}}~ {\left\| \bm{W} \left( \bm{A} \bm{y} - \bm{b} \right) \right\|}_2^2,~~\textup{s.t.}~~ \bm{y}^T \bm{D} \bm{y} + 2 \bm{f}^T \bm{y} = 0,
\end{align}
where $\bm{W} = \textup{diag} \left( \bm{w} \right)$ is a diagonal matrix with the elements of vector $\bm{w}$ on its main diagonal\footnote{It should be pointed out that the subscript $(\cdot)_{\left(k+1\right)}$ of $\bm{W}$ and $\bm{w}$ is dropped for notational simplicity.}, $\bm{w} = \left[ \sqrt{-\left[\bm{p}_{\left(k+1\right)}\right]_1}, \sqrt{-\left[\bm{p}_{\left(k+1\right)}\right]_2},..., \sqrt{-\left[\bm{p}_{\left(k+1\right)}\right]_L} \right]^T \in \mathbb{R}^L$,
\begin{align*}
\bm{A} &= \begin{bmatrix} -2 \bm{x}_1^T & 1 \\ \vdots & \vdots \\ -2 \bm{x}_L^T & 1 \end{bmatrix},~\bm{b} = \begin{bmatrix} r_1^2 - \left\| \bm{x}_1 \right\|_2^2 \\ \vdots \\ r_L^2 - \left\| \bm{x}_L \right\|_2^2 \end{bmatrix},\\
\bm{D} &= \begin{bmatrix} \bm{I}_d & \bm{0}_{d} \\ \bm{0}_{d}^T & 0 \end{bmatrix},~\bm{f} = \begin{bmatrix} \bm{0}_{d} \\ -1/2 \end{bmatrix},
\end{align*}
$\bm{0}_d \in \mathbb{R}^d$ denotes an all-zero vector of length $d$, and $\bm{I}_d \in \mathbb{R}^{d \times d}$ is the $d \times d$ identity matrix. Interestingly, the GTRS problem which aims to minimize a quadratic function subject to a single quadratic constraint, albeit usually nonconvex, possesses necessary and sufficient conditions of optimality from which effective algorithms can be derived \cite{ABeck}. To be specific, the exact solution of (\ref{GTRS}) is given by $\hat{\bm{y}}(\chi) = \left( \bm{A}^T \bm{W}^T \bm{W} \bm{A} + \chi \bm{D} \right)^{-1} \left( \bm{A}^T \bm{W}^T \bm{W} \bm{b} - \chi \bm{f} \right)$, where $\chi$ is the unique solution of $\psi(\chi) = \hat{\bm{y}}(\chi)^T \bm{D} \hat{\bm{y}}(\chi) + 2 \bm{f}^T \hat{\bm{y}}(\chi) = 0$ for $\chi \in I$, $I = \left( -\frac{1}{\chi_1 \left( \bm{D}, \bm{A}^T \bm{W}^T \bm{W} \bm{A} \right)}, \infty \right)$, and $\chi_1 \left( \bm{U},\bm{V} \right)$ denotes the largest eigenvalue of $\bm{V}^{-1/2} \bm{U} \bm{V}^{-1/2}$, given a positive definite matrix $\bm{V}$ and a symmetric matrix $\bm{U}$. Since $\psi \left( \chi \right)$ is strictly decreasing on $I$ (Theorem 5.2 in \cite{JJMore}), the optimal $\chi$ can be found using a simple bisection method.

So far, the two sub-problems in the AM procedure have been successfully addressed. We provide here a short remark on the convergence of our algorithm (termed SR-MCC by following the conventions in \cite{STomic2,STomic1,ABeck}). Analogous to Proposition 2 in \cite{XTYuan}, it can easily be deduced from (\ref{pp}), (\ref{px}), and the definitions of convex conjugate function that $\mathcal{A}_{\sigma}(\bm{x},\bm{p})$ increases at each AM step. Therefore, the sequence $\left\{ \mathcal{A}_{\sigma}\left(\bm{x}_{\left(k\right)},\bm{p}_{\left(k\right)}\right) \right\}_{k=1,2,...}$ generated by SR-MCC is non-decreasing. Based on the properties presented in \cite{WLiu}, one can further verify that $\mathcal{A}_{\sigma}\left(\bm{x}_{\left(k\right)},\bm{p}_{\left(k\right)}\right)$ is always bounded above. Then, convergence of the sequence to a limit point is assured.

The robustness of the MCC to a great extent hinges on the kernel size $\sigma$. In other words, a relatively small $\sigma$ assigns a much smaller weight (i.e., the role played by the auxiliary variable $p_i$) to the outliers during the iterations of HQ optimization, and hence achieves robustness against them. To ensure that the kernel size is always in the neighborhood of the best values \cite{WLiu}, we follow \cite{WLiu,JLiang} to adaptively select $\sigma$ at each HQ iteration based on the Silverman's heuristic \cite{WLiu,BWSilverman}, namely
\begin{equation}{\label{Silverman}}
\sigma_{\left(k+1\right)} = 1.06 \times \min\left\{ {\sigma_E}_{{\left(k+1\right)}}, R_{{\left(k+1\right)}}/1.34 \right\} \times L^{-1/5}, 
\end{equation}
where ${\sigma_E}_{{\left(k+1\right)}}$ is the standard deviation of the error $r_i^2 - {\|\bm{x}_{\left(k+1\right)} - \bm{x}_i\|}_2^2$ and $R$ is the error interquartile range \cite{WLiu}.

\begin{algorithm}[t]
	\SetAlgoNoLine
	\caption{SR-MCC for Robust TOA-Based Localization in NLOS Environments.}
	\KwIn{TOA-based range measurements $\{r_i\}$, sensor positions $\{\bm{x}_i\}$, and predefined $N_{\max}$, $K$, $\gamma$.}
	
	{
		
		$~~$\textbf{Initialize:} $\bm{x}_{\left( 0 \right)} = \bm{0}_d$ and $\sigma_{\left(0\right)} = \infty$.
		
		$~~$\textbf{for}$~k=0,1,\cdots~$\textbf{do}
		
			$~~~~$Update $\left\{ \bm{x}_{\left(k\right)}, \bm{p}_{\left(k\right)}, {\sigma}_{{\left(k\right)}} \right\}$ according to the AM steps in (\ref{eq:case2}) and kernel size updating rule in (\ref{Silverman}).
			
			$~~~~$\textbf{Stop} if predefined termination conditions are satisfied.
		
	    $~~$\textbf{end~with} $\tilde{\bm{x}} = \bm{x}_{\left( k+1 \right)}$.
		
	}
	\KwOut{Estimate of source location $\tilde{\bm{x}}$.}
\end{algorithm}

The termination criteria for the iterative algorithm SR-MCC are set as follows. The optimization variables $\bm{p}$ and $\bm{x}$ are iteratively updated until $k = N_{\max}$ or $\left\|\bm{x}_{\left( k+1 \right)} - \bm{x}_{\left( k \right)}\right\|_2 < \gamma$ is reached, where $N_{\max} \geq 1$ and $\gamma > 0$ are the predefined maximum number of iterations for the loop and tolerance parameter, respectively. For a clearer view, we summarize the whole procedure of SR-MCC in Algorithm 1.

\begin{table*}[!t]
	\renewcommand{\arraystretch}{0.6}
	\caption{Complexity of considered NLOS mitigation algorithms}
	\label{table_MCCTOA}
	\centering
	\begin{tabular}{c|c|c}
		\hline\hline
		\bfseries Algorithm & \bfseries Description & \bfseries Complexity\\
		\hline
		SR-MCC & Proposed MCC-based robust method & $\mathcal{O}(N_{\textup{HQ}}KL)$\\
		\hline
		SDP & SDP method in \cite{GWang} & $\mathcal{O}\left(L^{6.5}\right)$\\
		\hline
		SOCP & SOCP method in \cite{GWang} & $\mathcal{O}(L^{3.5})$\\
		\hline
		RSOCP & Robust SOCP method in \cite{SZhang} & $\mathcal{O}(L^{3.5})$\\
		\hline
		RMDSA & RMDSA method in \cite{WXiong} & $\mathcal{O}(N_{\textup{ADMM}}L^{2})$\\
		\hline
		SR-WLS & Bisection-based estimation method in \cite{STomic2} & $\mathcal{O}(KL)$\\
		\hline
		RSR-WLS & Bisection-based robust method in \cite{STomic1} & $\mathcal{O}(KL)$\\
		\hline\hline
	\end{tabular}
\end{table*}

It is not hard to find that the computational cost of operations in (\ref{eq:case2}a) is negligible compared to that in (\ref{eq:case2}b), i.e., in which the GTRS leading to a complexity of $\mathcal{O}(KL)$ \cite{STomic1} is incorporated. Here, $K$ is the number of steps taken by bisection search. The dominant complexity of our SR-MCC algorithm is thus $\mathcal{O}(N_{\textup{HQ}}KL)$, where $N_{\textup{HQ}}$ denotes the number of HQ iterations. In Table \ref{table_MCCTOA}, the computational complexity of SR-MCC is compared to several state-of-the-art approaches for TOA-based localization with NLOS mitigation\footnote{The complexity of the competitors has already been quantified in their respective studies and we simply list the results here. Interested readers are referred to the existing work \cite{STomic2,GWang,SZhang,STomic1,WXiong} for more details.}, where $N_{\textup{ADMM}}$ is the iteration number of the alternating direction method of multipliers in \cite{WXiong}. As our empirical results show, the proposed SR-MCC algorithm can already exhibit decent performance with a few number of $N_{\textup{HQ}}$ and $K$ and, hence, is fairly computationally simple. Note that we also provide comparison results in terms of average run-time in the next section for further confirmation.

\begin{figure*}[!t]
	\centering
	\subfigure[]{\includegraphics[width=2.36in]{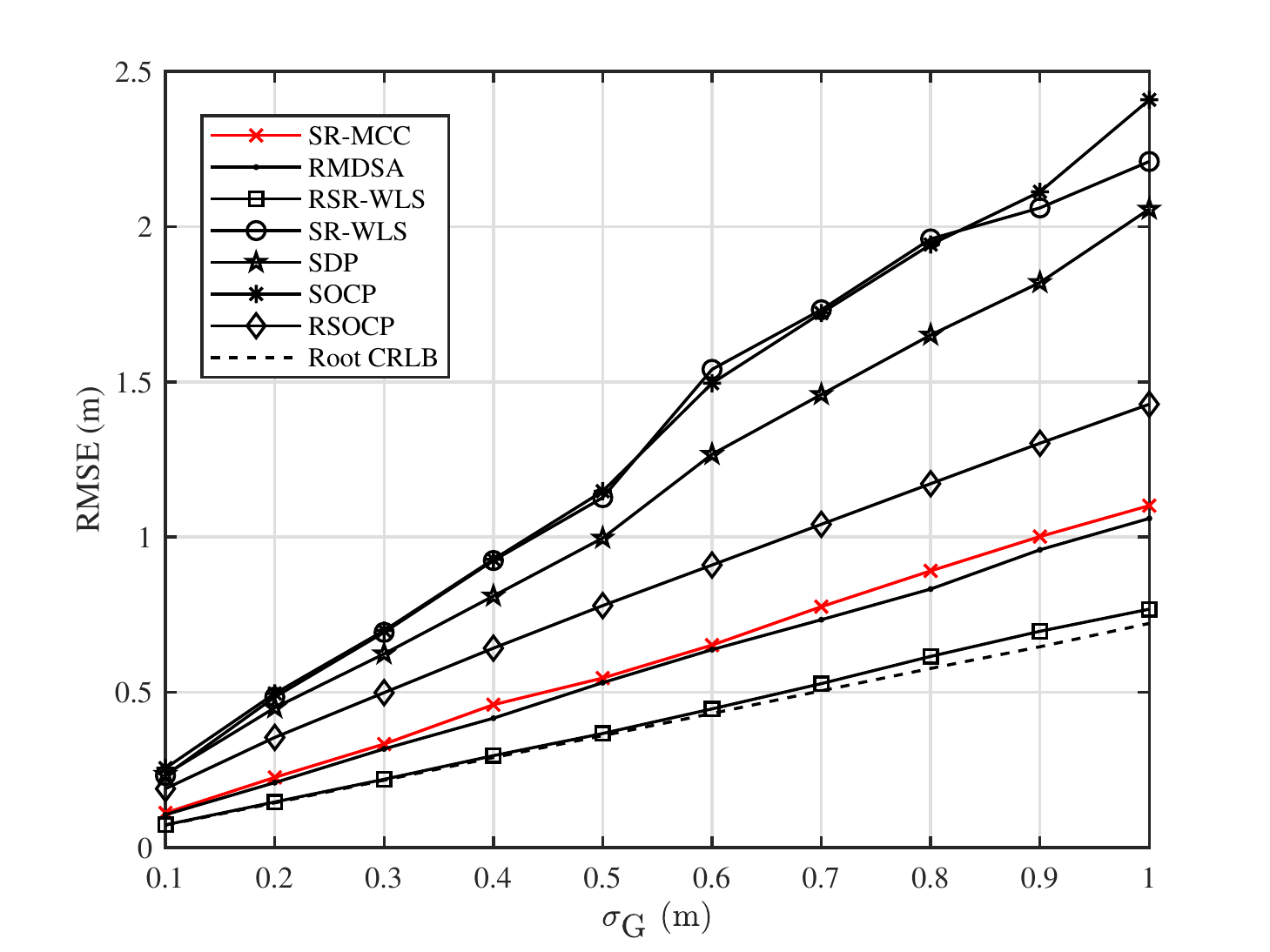}}%
	\subfigure[]{\includegraphics[width=2.36in]{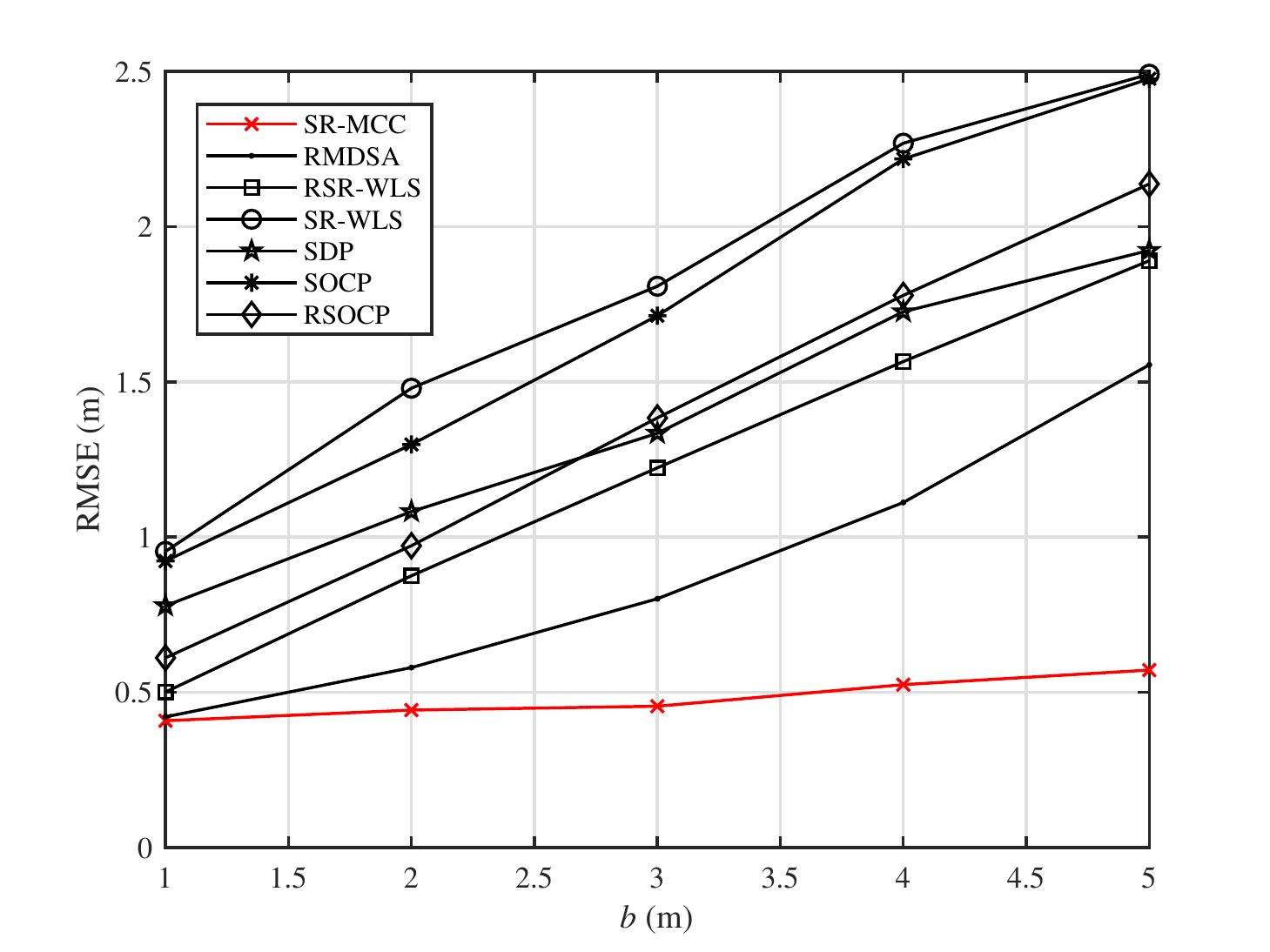}}
	\subfigure[]{\includegraphics[width=2.36in]{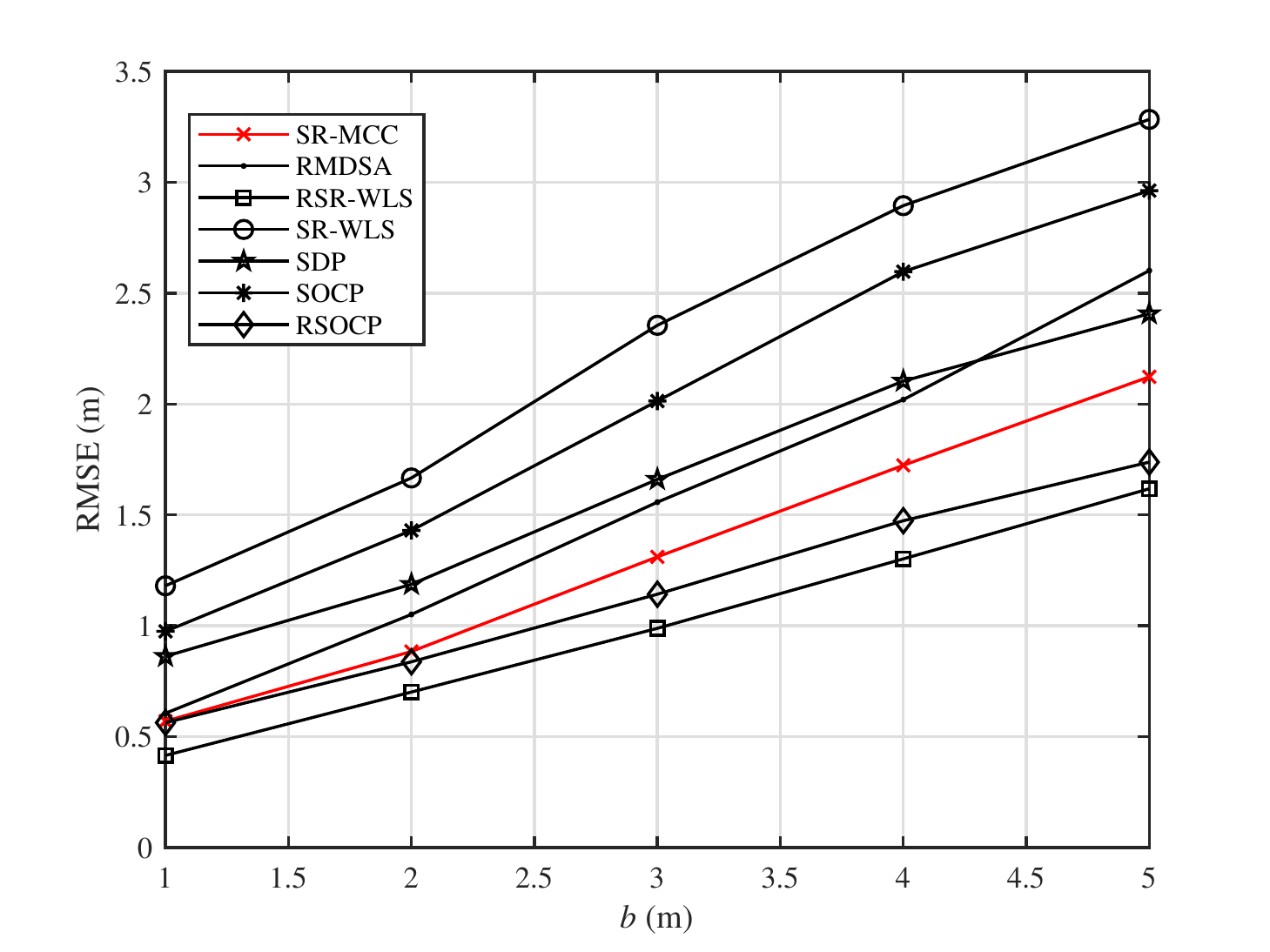}}
	\subfigure[]{\includegraphics[width=2.36in]{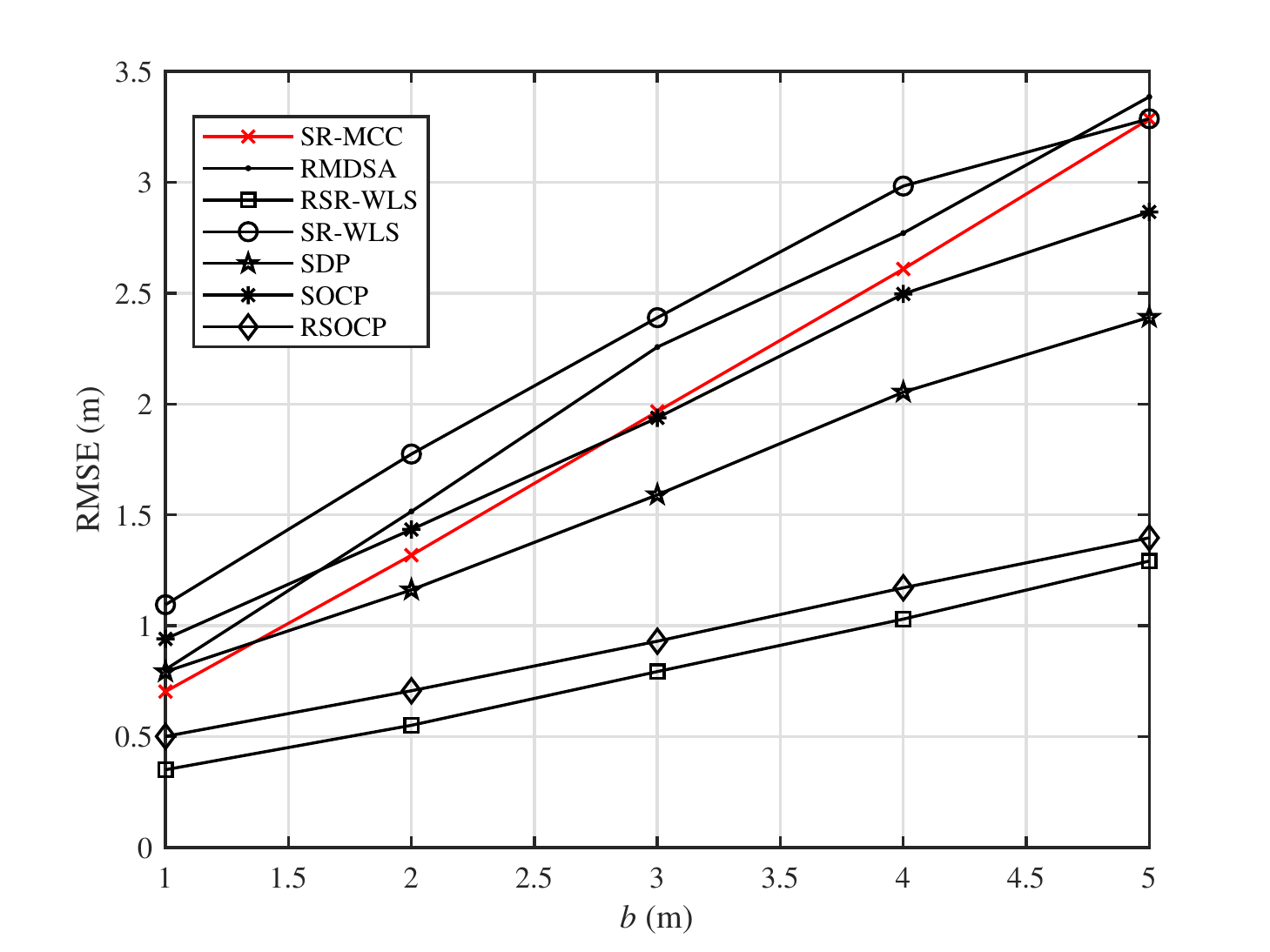}}
	\centering
	\caption{RMSE versus $\sigma_{\textup{G}}$ and $b$ in LOS and different NLOS scenarios, respectively. (a) $L_{\textup{NLOS}} = 0$. (b) $\sigma_{\textup{G}}^2 = 0.1, L_{\textup{NLOS}} = 2$. (c) $\sigma_{\textup{G}}^2 = 0.1, L_{\textup{NLOS}} = 5$. (d) $\sigma_{\textup{G}}^2 = 0.1, L_{\textup{NLOS}} = 8$.}
	\label{xiong1234}
\end{figure*}

\section{Numerical results}
\label{NR}
This section contains numerical investigations with the use of both synthetic and real experimental data. In addition to SR-MCC, state-of-the-art algorithms indicated in Table \ref{table_MCCTOA} are also included for comparison. We give a summary of the associated methods in Table \ref{table_MCCTOA2}, expatiating on the \textit{a priori} information required in their implementations. All the convex programs are realized using the CVX package \cite{MGrant}. Their infeasible runs are simply discarded\footnote{It is worth noting that our SR-MCC algorithm does not have this infeasibility problem.} and do not count towards the totals of Monte Carlo (MC) trials \cite{STomic2}. We set the stopping criteria of SR-MCC as $\gamma = 10^{-5}$, $N_{\textup{max}} = 10$, and $K=30$. On the other hand, algorithmic parameters of the existing methods remain unchanged as in their respective work. The computer simulations are all conducted on a Lenovo laptop with 16 GB memory and Intel i7-10710U processor.

\begin{table*}[!t]
	\renewcommand{\arraystretch}{0.6}
	\caption{Summary of methods incorporated in numerical investigations}
	\label{table_MCCTOA2}
	\centering
	\begin{tabular}{c|c}
		\hline\hline
		\bfseries Method & \bfseries Input\\
		\hline
		SR-MCC & Sensor positions and TOA-based range measurements\\
		\hline
		SDP & \tabincell{c}{Sensor positions, TOA-based range measurements,\\and noise variance}\\
		\hline
		SOCP & \tabincell{c}{Sensor positions, TOA-based range measurements,\\and noise variance}\\
		\hline
		RSOCP & \tabincell{c}{Sensor positions, TOA-based range measurements,\\noise variance, and upper bounds on NLOS errors}\\
		\hline
		RMDSA & Sensor positions and TOA-based range measurements\\
		\hline
		SR-WLS & Sensor positions and TOA-based range measurements\\
		\hline
		RSR-WLS & \tabincell{c}{Sensor positions, TOA-based range measurements,\\and upper bounds on NLOS errors}\\
		\hline\hline
	\end{tabular}
\end{table*}

\subsection{Results of synthetic data}

Basically, we consider a single-source localization setup with $L=10$ sensors and $d=2$. The source and sensors are all randomly deployed inside a 20 m $\times$ 20 m square region in each Monte Carlo (MC) run. In our setting, the Gaussian disturbance is assumed to be of identical variance $\sigma_{\textup{G}}^2$ for all choices of $i$s, and the NLOS bias is drawn from a uniform distribution on the interval $[0,b]$. Based on 3000 MC samples, the root mean square error (RMSE) defined as
\begin{equation}{\label{RMSE}}
\textup{RMSE} = \sqrt{\frac{1}{3000}\sum_{j=1}^{3000}{{\left\|\tilde{\mathbf{x}}^{\{j\}} - \mathbf{x}^{\{j\}}\right\|}^2}}
\end{equation}
is taken as the metric of positioning accuracy, where $\tilde{\mathbf{x}}^{\{j\}}$ denotes the estimate of source location ${\mathbf{x}}^{\{j\}}$ in the $j$th run.

We start with the ideal case, where all sensors are under LOS propagation (namely $L_{\textup{NLOS}} = 0$ with $L_{\textup{NLOS}}$ being the number of NLOS paths) and our mixture model of Gaussian and non-Gaussian distributions reduces to simply additive white Gaussian noise of variance $\sigma_{\textup{G}}^2$. Fig. \ref{xiong1234} (a) plots the RMSE versus $\sigma_{\textup{G}}^2$ for all the considered algorithms in this scenario, with the Cram\'{e}r-Rao lower bound (CRLB) \cite{HCSo} being included for benchmarking purposes. It is observed that SR-MCC, RMDSA, and RSR-WLS have much lower RMSEs than the others, though SR-MCC is slightly inferior to RMDSA and RSR-WLS. Among all the methods, only the solution accuracy of RSR-WLS can achieve the CRLB up to low Gaussian noise levels. Fixing the variance of noise as $\sigma_{\textup{G}}^2 = 0.1$, Figs. \ref{xiong1234} (b), \ref{xiong1234} (c), and \ref{xiong1234} (d) subsequently compare the performances of diverse approaches under three different and typical NLOS conditions. We clearly see from Fig. \ref{xiong1234} (b) that SR-MCC outperforms the other methods for all $b$s in a mild NLOS environment with $L_{\textup{NLOS}} = 2$. As depicted in Fig. \ref{xiong1234} (c), when the number of NLOS connections is moderate, i.e., $L_{\textup{NLOS}} = 5$, our proposed scheme is superior to RMDSA, SR-WLS, SDP, and SOCP while yielding a bit higher RMSE values than RSR-WLS and RSOCP. Fig. \ref{xiong1234} (d) illustrates the RMSE versus $b$ in an extremely dense NLOS environment with $L_{\textup{NLOS}} = 8$. Although SR-MCC degrades in a sense that it cannot overwhelmingly outperform SOCP and SDP in this case, it still produces the minimum RMSE for all $b$s among SR-MCC, RMDSA, and SR-WLS, which are the only schemes whose operations require no more than the sensor locations and TOA-based distance measurements. On the contrary, the other solutions more or less take advantage of and are reliant upon additional \textit{a priori} knowledge of the noise variance and/or error bound. Apart from these, the performances of all the considered algorithms deteriorate as $\sigma_{\textup{G}}$ or $b$ grows.

\begin{figure}[!t]
	\centering
	\includegraphics[width=3.5in]{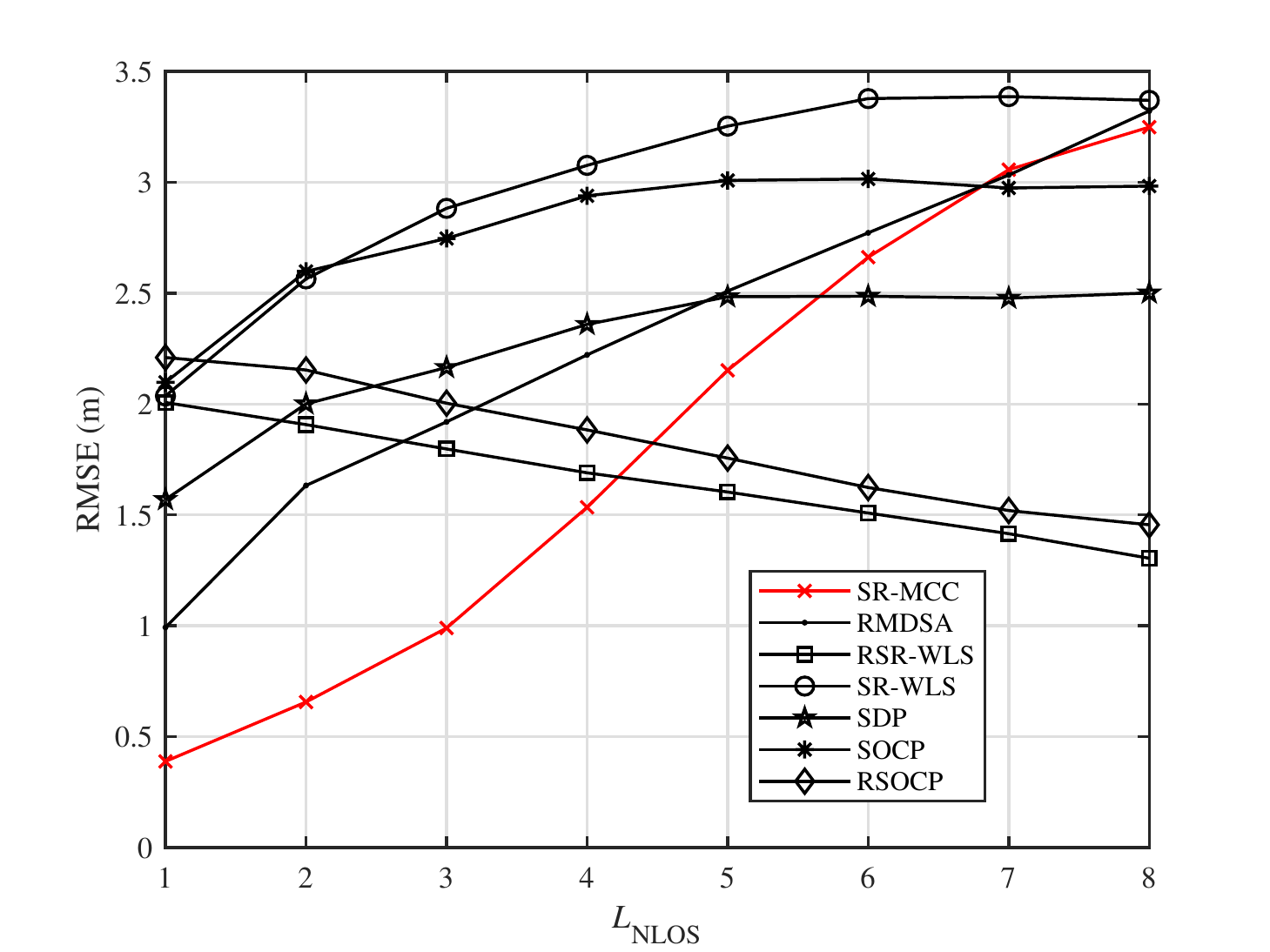}%
	\caption{RMSE versus $L_{\textup{NLOS}}$ at $\sigma_{\textup{G}}^2 = 0.1$ and $b = 5$.}
	\label{xiong5}
\end{figure}

To summarize, it is preferred to employ our SR-MCC method if the number of the NLOS connections is not large enough. This actually coincides with the properties of the correntropy measure counted on in building our objective function (see Section \ref{PFP}), and is further verified in Fig. \ref{xiong5} demonstrating the RMSE versus $L_{\textup{NLOS}} \in \left[ 1, 8 \right]$ at $\sigma_{\textup{G}}^2 = 0.1$ and $b = 5$. Apart from the statistical robustness of the Welsch loss to large errors as showcased in Fig. \ref{Losses_Comp}, more explanations for the outstanding performance of the MCC-based robustification strategy in several mixed LOS/NLOS environments are given below from the perspective of HQ iterations. As the iteration summarized in Algorithm 1 proceeds, the auxiliary variables in $\bm{p}$ updated according to (\ref{s_subp1}) play the role of Gaussian-like weighting functions \cite{WLiu}, thus capable of mitigating the adverse effects of large SR fitting errors in the GTRS (\ref{GTRS}) to a great extent \cite{JLiang}.

\subsection{Results of real experimental data}

\begin{figure*}[!t]
	\centering
	\subfigure[]{\includegraphics[width=2.8in]{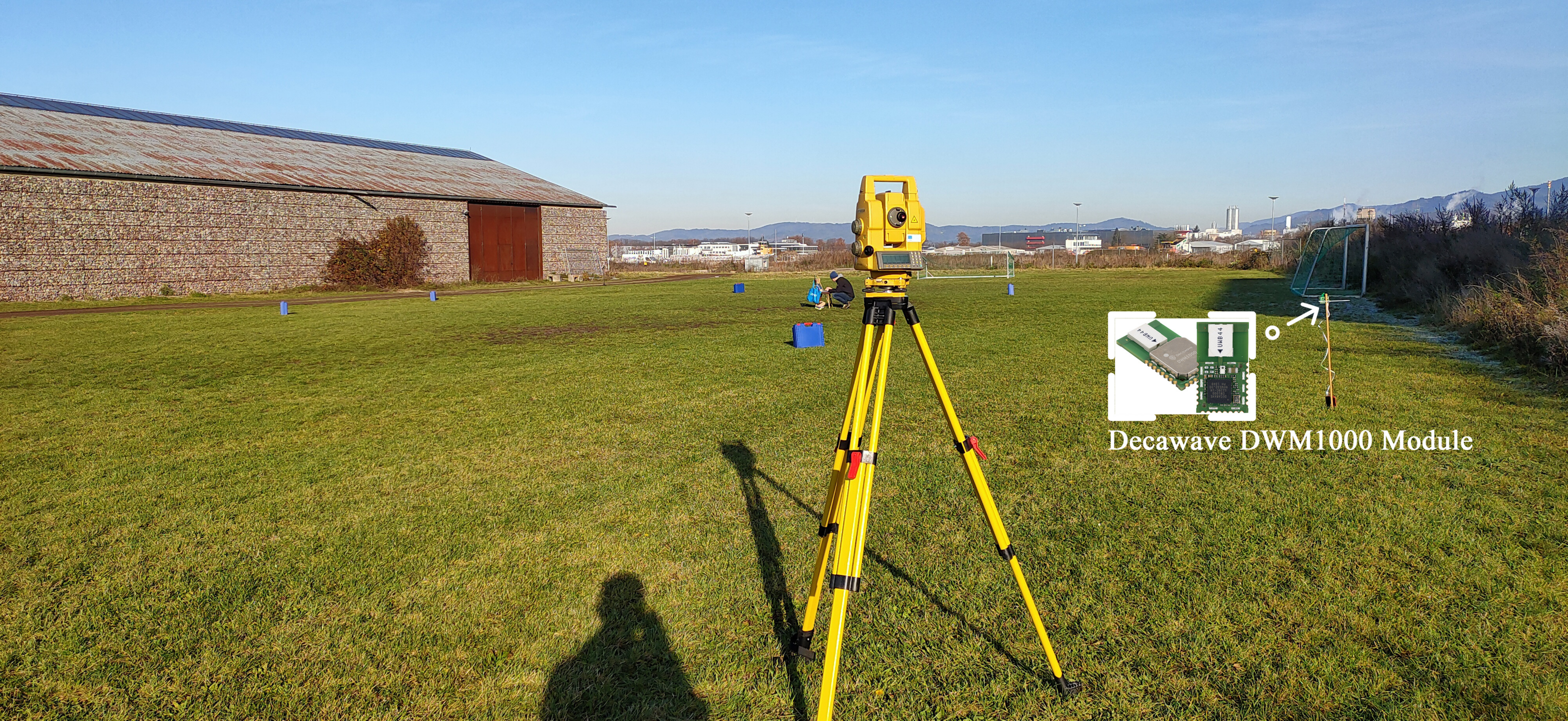}}%
	\subfigure[]{\includegraphics[width=1.8in]{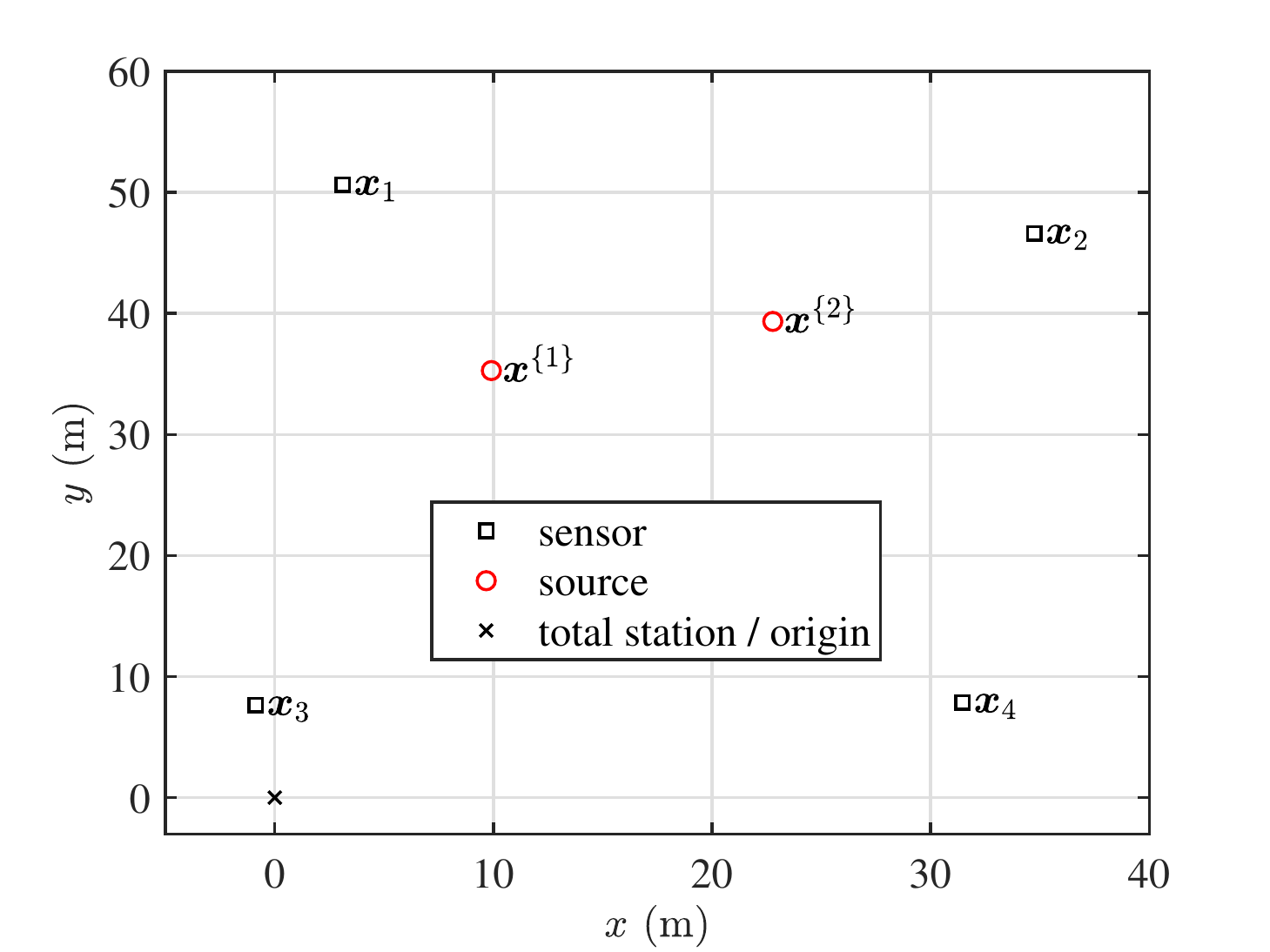}}
	\centering
	\caption{Experimental environment for data collection. (a) Real-world deployment. (b) 2-D illustration of localization geometry.}
	\label{Exp_out}
\end{figure*}

This subsection substantiates the efficacy of SR-MCC through the use of real experimental data. The localization experiments have been conducted within a 50 m $\times$ 50 m open area (see Fig. \ref{Exp_out}) at the Technische Fakult\"at campus of the University of Freiburg, Freiburg im Breisgau, Germany, and the data have been acquired by using the ranging systems developed based on Decawave DWM1000 modules \cite{ARJRuiz,Deca}. Each DWM1000 module is an IEEE 802.15.4-2011 UWB implementation based on Decawave's DW1000 UWB transceiver integrated circuit \cite{Deca}, and we have installed five modules in our real-world experiments. Among them, four modules attached to the wooden rods with know positions (see Fig. \ref{Exp_out}(a)) are specified as the sensors, whereas the remaining one serves as the source to be located. The power is supplied using the power banks. For the purpose of testing, two reference points are considered, and the source stops its movements and stays long enough at each of the reference points, such that 100 sets of steady two-way ranging measurements between the source and sensors are performed. By deploying a Topcon GPT-8203A total station at the origin, we set up the coordinate system (shown in Fig. \ref{Exp_out}(b)) and the true positions of the sensors and reference points can be measured. Here, we have $d = 2$ because the source and all the sensors are intentionally always of the same height 1.2 m. The positions of the sensors and reference points are tabulated in Table \ref{table_MCCTOA3}. In particular, several obstructions are created in the path between the source and and first sensor on purpose to construct the NLOS environments.

\begin{table*}[!t]
	\renewcommand{\arraystretch}{0.6}
	\caption{Sensor and reference point positions}
	\label{table_MCCTOA3}
	\centering
	\begin{tabular}{c|c|c}
		\hline\hline
		\bfseries Attribute & x (m) & y (m) \\
		\hline
		 $1$st sensor & 3.1068 & 50.6350\\
		 \hline
		 $2$nd sensor & 34.7464 & 46.6166\\
		 \hline
		 $3$rd sensor & -0.8732 & 7.6484\\
		 \hline
		 $4$th sensor & 31.4618 & 7.8664\\
		 \hline
		 $1$st ref. point & 9.9064 & 35.2822\\
		 \hline
		 $2$th ref. point & 22.7794 & 39.3434\\
		\hline\hline
	\end{tabular}
\end{table*}

\begin{figure}[!t]
	\centering
	\includegraphics[width=3.5in]{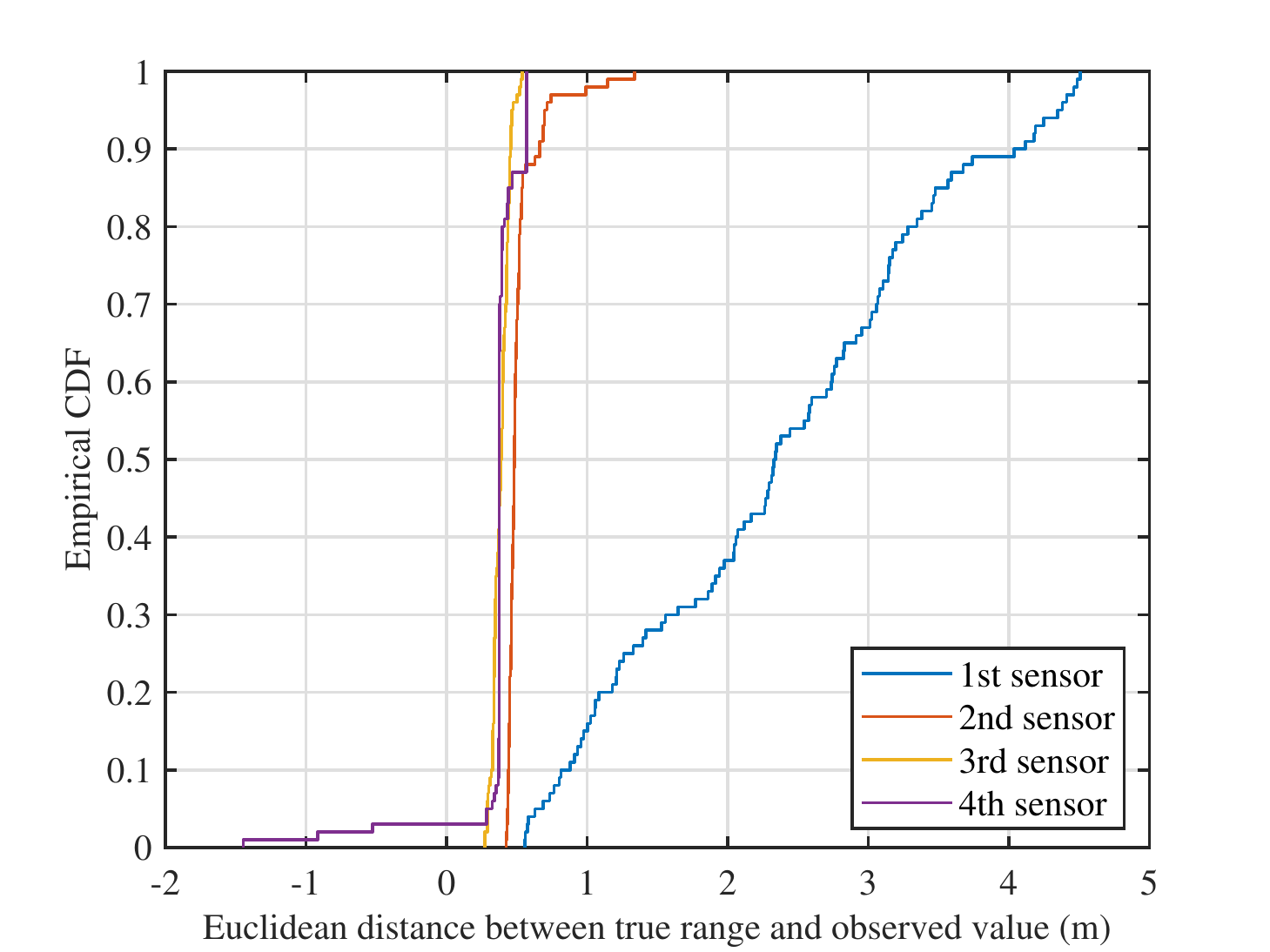}%
	\caption{Empirical CDF of Euclidean distance between true range and observed value based on 50 data sets acquired at 2 reference points.}
	\label{xiongrealCDF}
\end{figure}

To determine the upper bound $\bar{b}$ on the NLOS errors needed by RSOCP and RSR-WLS, Fig. \ref{xiongrealCDF} plots the empirical cumulative distribution function (CDF) of the Euclidean distance between the range measurement and its true value. Following the similar strategy to \cite{HChen2}, we set it as $\bar{b} = 4$ associated with the probability of $90\%$ in Fig. \ref{xiongrealCDF}. Furthermore, the noise variance required by SDP, SOCP, and RSOCP is set as $\sigma_{\textup{G}}^2 = 0.02$. Table \ref{table_MCCTOA4} shows the average run-time recorded using MATLAB commands \texttt{tic} and \texttt{toc} and RMSE\footnote{The number of samples in the original definition of RMSE in (\ref{RMSE}) is changed accordingly.} values for different algorithms. The results of the measured elapsed time roughly accord with the complexity analysis in Table \ref{table_MCCTOA}. We see that the amounts of average run-time for the SOCP/SDP-based approaches all exceed 1 s, reinforcing the general consensus that convex optimization usually results in non-negligible computational overheads. In contrast, SR-MCC, RMDSA, SR-WLS, and RSR-WLS are computationally much simpler. We point out that the complexity level of SR-MCC is a bit higher than RMDSA, SR-WLS, and RSR-WLS, as it involves solving a series of GTRSs. Nonetheless, our SR-MCC method has the best localization accuracy in terms of the RMSE.

\begin{table*}[!t]
	\renewcommand{\arraystretch}{0.6}
	\caption{Performance comparison using real experimental data}
	\label{table_MCCTOA4}
	\centering
	\begin{tabular}{c|c|c}
		\hline\hline
		\bfseries Algorithm & \bfseries Run-Time \mdseries (s) & \bfseries RMSE \mdseries (m) \\
		\hline
		SR-MCC & 0.0172 & 0.564 \\
		\hline
		SDP & 1.2784 & 1.246 \\
		\hline
		SOCP & 1.3555 & 1.284 \\
		\hline
		RSOCP & 1.3886 & 1.670 \\
		\hline
		RMDSA & 0.0014 & 1.327 \\
		\hline
		SR-WLS & 0.0072 & 1.451 \\
		\hline
		RSR-WLS & 0.0034 & 1.489 \\
		\hline\hline
	\end{tabular}
\end{table*}

\section{Conclusion}
\label{CC}
In this paper, we have devised a novel NLOS mitigation technique for TOA-based source localization. Our key idea is to utilize the correntropy-based error measure to achieve robustness against the bias-like NLOS errors. An HQ framework has been adopted to deal with the nonlinear and nonconvex correntropy-induced optimization problem in a computationally inexpensive AM fashion. The mentionable merit of the proposed algorithm is its low prior knowledge requirement. Extensive numerical results have confirmed that our method can outperform several existing schemes in terms of localization accuracy, especially in mixed LOS/NLOS environments where the number of NLOS connections $L_{\textup{NLOS}}$ is not large enough. Nevertheless, the presented approach has its limitation that it might suffer from the loss of localization accuracy as $L_{\textup{NLOS}}$ increases. An important direction for the future work is to further robustify the estimator w.r.t. $L_{\textup{NLOS}}$, and a possible solution can be combining the statistical robustification scheme with the worst-case criterion.


%
\section*{Data availability}
The datasets generated during the current study are not publicly available due to the simplicity of reproduction but are available from the corresponding author on reasonable request.

\section*{Conflict of interest}
The authors declare that they have no conflict of interest.



\end{document}